\documentstyle[aps,preprint,prl]{revtex}

\begin{document}
{\tighten
\preprint{\vbox{\hbox{UCSD/PTH 94--13}}}

\title{QCD Flux Tubes as Sigma Model Relics}

\author{Katherine M.~Benson,
 Aneesh V.~Manohar, and Maha Saadi}
\address{Department of Physics\\
University of California,~San Diego\\  La Jolla, California 92093}

\bigskip
\date{revised November 22, 1994}

\maketitle
\begin{abstract}

  We describe flux tubes and their interactions in a low energy sigma
  model induced by $SU({N_F}) \rightarrow SO({N_F})$ flavor symmetry
  breaking in $SO(N_c)$ QCD.  Gauge confinement manifests itself in
  the low energy theory through flux tube interactions with unscreened
  sources. The flux tubes mediating confinement act as Alice strings
  in their cores, a phenomenon which may occur for $\pi_2$-line
  defects in physically realized systems.
\end{abstract}

}% end the tighten

\narrowtext

Nonlinear sigma models successfully describe many low energy QCD
phenomena. However, they have not captured the hallmark feature of
QCD: confinement, where potentials between $q\bar{q}$ pairs grow
linearly with separation.  Such linear potentials arise from flux
tubes between unscreened $q$ and $\bar q$ sources for pure Yang-Mills
QCD. Conventional sigma models describe the low-energy dynamics of QCD
due to the global flavor symmetry breaking $SU({N_F})_L \times SU({N_F})_R
\rightarrow SU({N_F})_{diag}.$ Global flux tubes do not occur in these
models, as the vacuum manifold $G/H$ forms a Lie group, with trivial
$\pi_2$. This absence of flux tubes is consistent with the absence of
unscreened sources in QCD with fundamental quarks.

Witten \cite{wittenon} noted that an $SO(N_c)$ gauge theory of QCD
induces a different sigma model, whose topology {\em can} support flux
tubes. The theory has ${N_F}$ left-handed quarks which transform as
(real) fundamentals of $SO(N_c)$.  The color singlet condensate
$\langle q_{Li}^\alpha \,\,q_{Lj}^\alpha\rangle$ (where $\alpha$ and
$i$ are color and flavor indices respectively) breaks the $SU({N_F})$
flavor symmetry to its $SO({N_F})$ subgroup. There is a baryon number
$SO(N_c)^2$ anomaly which leaves an anomaly free conserved $Z_2$
baryon number. The Goldstone modes are described by a $SU({N_F}) /
SO({N_F})$ nonlinear sigma model, with skyrmions ($\pi_3(G/H)$ = $Z_2$,
for ${N_F} \ge 4$) and flux tubes ($\pi_2(G/H)$ = $Z_2$, for ${N_F} \ge
3$). This defect classification makes sense: (a) baryons are
identified with antibaryons since real quarks are identified with
antiquarks; (b) there are $Z_2$ flux tubes which confine spinor
sources, since only an even number of spinors can be screened by
fundamental quarks.

In this letter we construct the flux tubes in this theory and show
that their interactions with skyrmions and spinor sources obey
heuristic expectations.  We show that confinement in an $SO(N_c)$
gauge theory can manifest itself in the low energy sigma model through
flux tube interactions with unscreened sources. We proceed as follows:
we derive the unique flux tube form with minimal energy; examine its
classical stability and dynamics; and then study its quantum
stability, spectrum, and interactions. We discuss how flux tubes act as
Alice strings in their cores, form loops that support skyrmion number,
and mediate the confinement of spinor sources. We skip
many technical details, which appear in %a longer paper by two of us~
\cite{longskyrme}.

\centerline{{\bf Finding Nontrivial Flux Tubes}}

The sigma model for $SO(N_c)$ QCD describes the dynamics of a sigma
field $\Sigma$, which encodes the orientation of the fermion
condensate $\langle q^\alpha_{Li} \,\, q^\alpha_{Lj}\rangle$ with respect
to its standard form $\propto \delta_{ij}$, with
$\Sigma=1$. Under an $SU({N_F})$ transformation by $g$, $\Sigma$
transforms as $\Sigma \rightarrow g\Sigma g^T$, so that a generic
$\Sigma$ has the form $gg^T$ for some group element $g$.
$\Sigma={ {\mbox{1\hskip-0.22em\relax l}}\,}$ is invariant under a $g$
transformation if $g\in
SO({N_F})$, so $\Sigma$ is an element of the coset space
$G/H=SU({N_F})/SO({N_F})$, which is not a Lie group.

$G$ invariance and cylindrical symmetry fixes the
form of the minimal energy flux tube to be $\Sigma =
g(\theta) \Sigma_o(r){{g}^T} (\theta)$ --- an $r$-dependent
vev, with angle-dependent group rotation, where $(r,\theta,z)$ are
cylindrical coordinates. Choosing $\Sigma_o = { {\mbox{1\hskip-0.22em\relax
l}}\,} $ at spatial
infinity forces $g(\theta)$ to lie in $H  = SO({N_F})$, and to commute
with $\Sigma_o(r=0),$ for a nonsingular solution with finite energy.

$\Sigma$ can be further specified using basis generators of the
$su({N_F})$ Lie Algebra, with Cartan norm $\mbox{tr}\, T_a T_b =
{\textstyle\frac{1}{2}}
\delta_{ab}$. The rank ${N_F}-1$ Cartan subalgebra has as a basis
diagonal matrices $T_d$,
\begin{equation}
\label{defTd}
T_d = ( 2d(d +1)\, )^{-1/2}\ {\rm diag} (1,\ldots,1, -d,0,
\ldots,0),
\end{equation}
with ones in the first $d$ entries and $1\le d\le{N_F} -1$. We can
overspecify them, sacrificing the Cartan norm, by Pauli matrices
$\{ {\textstyle\frac{1}{2}}\tau_{z\,(jk)\,}\}$ in all $(jk)$ subplanes of the
${N_F}$
dimensional vector space. Off-diagonal generators $\{
{\textstyle\frac{1}{2}}\tau_{x\,(jk)\,}$,
${\textstyle\frac{1}{2}}\tau_{y\,(jk)\,}\}$ complete the basis.
The antisymmetric matrices $\{T_h\} \equiv\{{\textstyle\frac{1}{2}}
\tau_{y\,(jk)\,}\}$
generate the unbroken $H = SO({N_F})$ symmetry, while the symmetric
matrices $\{T_b\} \equiv\{ {\textstyle\frac{1}{2}}\tau_{x\,(jk)\,},
{\textstyle\frac{1}{2}}\tau_{z\,(jk)\,} \}$ are broken generators.
$\Sigma_o(r)$, a unitary symmetric matrix, can be written as $\exp
{\{i F_b(r) T_b\} }$. Thus $\Sigma$ assumes the most general
nonsingular form $\Sigma = h(\theta) \exp {\{ i F_b(r) T_b)\} }
h^{-1}(\theta),$ for $h(\theta) \in H$ and $F_b(r)$ ranging from zero
at infinity to $ 2\pi n \,\,\delta_{bb'}$ (for some direction
$b'$) at the origin.

We construct a non-trivial flux tube $\Sigma$ from the exact
sequence
\[
\pi_2\left( SU({N_F})\right) = 0 \rightarrow \pi_2\left(
SU({N_F})/SO({N_F})\right)
\rightarrow
\pi_1\left( SO({N_F})\right) = Z_2 \rightarrow
\pi_1\left( SU({N_F})\right) = 0 .
\]
That is, $g g^T$ gives a nontrivial $\Sigma$ only if $g$ corresponds
to some mapping from the plane to $SU({N_F}),$ with boundary values in
the $SO({N_F})$ subgroup. Furthermore, when parametrized as a family
of loops, $g$ must start at the identity and end on a nontrivial loop
in $SO({N_F})$. Taking $\left (\alpha\in[0,2\pi],
\beta\in[0,\pi]\right) $ as our coordinates on the plane, these
criteria become
\begin{equation}
\label{gbc}
g(\alpha, \beta) = \left\{
\begin{array}{ll}
{ {\mbox{1\hskip-0.22em\relax l}}\,} & \mbox{when $\alpha= 0$, $\alpha=2\pi$,
\ or $\beta = 0$}\\
h^2(\alpha) & \mbox{when $\beta = \pi$},
\end{array}
\right.
\end{equation}
where $h^2(\alpha)$ is a nontrivial loop in $SO({N_F}).$

Through technical arguments, two of us show that such
trivializations of $SO({N_F})$ loops in $SU({N_F})$ have minimal energy
only when they induce $\Sigma$ of a very limited form \cite{longskyrme}.
To minimize energy, $h^2(\alpha)$ is geodesic:
$h(\alpha) = \exp{\{i\alpha \,\, T_{h\Box}\}}$, where $T_{h\Box} =
\{{\textstyle\frac{1}{2}} \tau_{y\,}\}$ in some plane $\Box $ --- taken here as
$(12)$
for concreteness. This geodesic loop determines the
associated flux tube
$\Sigma$:
\begin{equation}\begin{array}{c}
\label{Sigfinform} \Sigma(r, \theta)=h(\theta) \ b(r) \
h^{-1}(\theta),\ \ \mbox{with }\\[3pt]
h(\theta)= \exp{\{i \theta
\,\, T_{h\Box}\}},\ b(r) =  \exp{\{i F_d(r) T_d\, \}},\\[3pt]
\mbox{and } F_1(r=0) = 2\pi; \ \ F_d(r\rightarrow\infty) = 0.
\end{array}\end{equation}
Here $T_{h\Box}$
generates the loop $h^2(\alpha)$, and $T_d$ are the Cartan
generators from eq.~(\ref{defTd}).  The boundary conditions on $F_d$ stem from
eq.~(\ref{gbc}), requiring $g(\alpha,\beta)$ to trivialize
$h^2(\alpha)$.

Note that when $F_{d>1}$ vanishes, $\Sigma$ lies entirely within a
planar $SU(2)$ subgroup of $SU({N_F}),$ as $T_1 = \tau_{z\,\Box\,}/2$.
This produces the simple form
\begin{equation}
\label{Sigrexp}
\Sigma_\Box(r, \theta) =
{ {\mbox{1\hskip-0.22em\relax l}}\,} + \left(\,\,\cos (F/2) - 1 \,\,\right)\ {
{\mbox{1\hskip-0.22em\relax l}}\,}_\Box
+ i \sin{(F/2 )} \left(\,\, \cos{\theta}\ \tau_{z\Box} -
\sin{\theta}\ \tau_{x\Box} \,\,\right).
\end{equation}
where $F\equiv F_1$, ${ {\mbox{1\hskip-0.22em\relax l}}\,}_\Box$ gives
the identity in the plane $\Box$ and vanishes outside it, and ${
  {\mbox{1\hskip-0.22em\relax l}}\,}$ is the usual $SU({N_F})$
identity. In this letter, we consider only such planar flux tubes,
having shown in \cite{longskyrme} that they minimize energy for the
action discussed here.

\centerline{{\bf Flux Tube Stability and Dynamics}}

As a minimal model for $\Sigma$ with stable skyrmions, consider the
Skyrme lagrangian \cite{anw} \footnote{We use QCD sigma model
  parameters $e, F_\pi$, $m_\pi$, etc., although these are not the
  usual pions. There is also a Wess-Zumino term, which vanishes for
  constant-$z$ flux tubes.}
\begin{equation}
\label{SSkyrme}
%S_o= \int\,\, 2\pi rdr\,d\theta \, dz\, dt\ \left \{\,\,        \right \}
{\cal L}_0= \frac{F_\pi^2}{16} \,\,\mbox{tr}\, \,{\partial}_\mu \Sigma
{\partial}^\mu {\Sigma}^\dagger
        +\frac{1}{32e^2} \,\,\mbox{tr}\,\,
  {\hbox{\bf [}  {\Sigma }^\dagger {\partial}_\mu \Sigma,{\Sigma }^\dagger
{\partial}_\nu \Sigma  \hbox{\bf ]}}^2
\ .
\end{equation}
This model does not have stable flux tubes,
for under the rescaling $\Sigma(r,\theta,z) \rightarrow \Sigma (\lambda
r,\theta,z)$, the tension of a finite flux tube can decrease.
Flux tubes diffuse to infinite size ($\lambda = 0$) to lower their energy.

There are several modifications to the minimal ${\cal L}_0$, or to the
minimal form for $\Sigma$, that lead to stable flux tubes
\cite{longskyrme}. One natural choice is to give non-zero mass to the
quarks in the original QCD theory.  This induces a mass term in the
effective lagrangian,
\begin{equation}
\label{SMass}
%S_m = \int\,\, 2\pirdr\,d\theta \,dz\, dt\
{\cal L}_m =       \frac{F_\pi^2 \, m_\pi^2}{16} \mbox{tr}\,
        \left(  \Sigma  + {\Sigma}^\dagger - 2\cdot {
{\mbox{1\hskip-0.22em\relax l}}\,} \right)
\ ,
\end{equation}
which gives mass to the Goldstone bosons, and also stabilizes the
flux tube solutions.

The Skyrme action eq.~(\ref{SSkyrme}) gives gradient energy density to
the flux tube:
\begin{equation}
\label{Egradgen}
%E_{0}= \frac{\pi F_\pi^2}{8} \int\,\, rdrdz\
\rho_{0}= \frac{F_\pi^2}{16}
        \mbox{tr}\, \left \{
        \left(\,F'_d\, T_d\, \right)^2
        +\frac{1}{r^2} \,\, \left( {\tilde{T}}^2 -
        {\hbox{\bf [} \,F'_d\, T_d\, ,\tilde{T} \hbox{\bf ]}}^2 \right)
        \right \} \ ,
\end{equation}
where
\begin{equation}
\label{defTtilde}
{\tilde{T}} \equiv
b^{-1} (r) \ {\hbox{\bf [} \ T_{h\Box}\ ,\ b(r)\  \hbox{\bf ]}}
\ ,
\end{equation}
and $r$ has been rescaled to dimensionless units
$eF_\pi \, r_{phys}$. For the planar solution eq.~(\ref{Sigrexp}),
this gives the energy density
\begin{equation}
\label{Egrad}
%E_{0}= \frac{\pi F_\pi^2}{8} \int\,\, rdrdz\,\
\rho_{0}= \frac{F_\pi^2}{16}
        \left \{ {\textstyle\frac{1}{2}}  {F'} ^2
        +\frac{1}{r^2} \left( 1- \cos F \right) \left( 1 + {F'}^2 \right)
        \right \} \ .
\end{equation}
To this term we add potential energy from the pion mass term
eq.~(\ref{SMass}), % to give the total energy density
\begin{equation}
\label{Eflux}
%E= \frac{\pi F_\pi^2}{8} \int\,\, rdrdz\,\
%\rho= \frac{F_\pi^2}{16}
%        \left \{ {\textstyle\frac{1}{2}} {F'} ^2
%        +\frac{1}{r^2} \left( 1- \cos F \right) \left( 1 + {F'}^2 \right) +
%        \right \} \ ,
\rho_m =  \frac{\lambda^2 F_\pi^2}{16}
        \left( 1 - \cos (F/2)\right)\ ,
\end{equation}
where $\lambda = 2 m_\pi/e\,F_\pi.$ Together, these energy
contributions determine a nonlinear equation of motion for $F$, which
we solve numerically. Solutions appear in Figure 1 for different
values of $\lambda$, including the physical $\lambda_0=0.236$
($e=2\pi$, $m_{\pi}=138$ Mev and $F_{\pi}=186$ Mev). Increasing
$\lambda$ raises the flux tube's energy density while shrinking its
core size.  Inside the core, $F$ falls linearly from $F(0) = 2\pi$;
outside, it scales as the hyperbolic Bessel function $F \sim
x^{-1/2}\,\exp{\{-x\}},$ with $x = \lambda r/2$.  $2\lambda^{-1}$ sets
the flux tube's dimensionless core size, which gives physical core
size $m_\pi^{-1}$. The flux tube tension is proportional to $F_\pi^2$
--- which is proportional to the number of colors $N_c$ in the large
$N_c$ limit.  This supports its identification with the confining
force between spinor sources, since the Casimir for $SO(N_c)$ spinors
is of order $N_c$. Numerically we find tension $4.6 F_\pi^2$ when
$\lambda = \lambda_0$, scaling linearly with $\lambda$.

\centerline{{\bf Quantum Stability and Spectrum}}

The quantum numbers and low-lying internal excitations of the flux tube are
given by quantizing the zero-mode collective coordinates \cite{Balach},
\begin{equation}
\label{defSigt}
\Sigma(t, r, \theta) = A(t) \ \Sigma(r,\theta)\ A^{-1} (t) \ ,
\end{equation}
where $A(t)$ rotates about $n_h T_h$ with dimensionless frequency
$\omega$.~\footnote{We neglect other excitations, such as bending
  modes.} These modes have rotational energy confined to the string
core, which can be calculated from the Skyrme action eq.~(\ref{SSkyrme}):
\begin{equation}
\label{Ezms}
%E_{\omega} = \frac{\pi F_\pi^2 \,\,\omega^2}{8} \int\,\, rdrd\theta dz\
\rho_{\omega} = \frac{F_\pi^2 \,\,\omega^2}{16}
\mbox{tr}\, \left ( \hat{T}^2 -
        {\hbox{\bf [} \,F'_d\, T_d\, ,\hat{T} \hbox{\bf ]}}^2 -
        \frac{1}{r^2}   {\hbox{\bf [} \,{\tilde{T}}\, ,\hat{T} \hbox{\bf ]}}^2
        \right) \ .
\end{equation}
with ${\tilde{T}}$ from eq.~(\ref{defTtilde}) and
\begin{equation}
\label{defThat}
\hat{T}  \equiv
b^{-1}(r) \ {\hbox{\bf [} \ h^{-1} (\theta)\ n_h T_h\ h (\theta)\ ,\ b(r)\
\hbox{\bf ]}} \ .
\end{equation}
Calculating eq.~(\ref{Ezms}) is tedious, and described in
\cite{longskyrme}.  We show there that the planar vacuum
survives quantum fluctuations due to zero modes.  It has the
classical two dimensional Lagrange density
\begin{equation}
\label{MQSkyrme}
L({\tilde z},t) = -\tilde{\rho}
-\frac{\Lambda_{{\scriptscriptstyle\Box}}}{2}
(\mbox{tr}\,{A}^\dagger{\dot A}T_{h{\scriptscriptstyle\Box}})^2
 -\frac{\Lambda_{{\scriptscriptstyle\Box'}}}{2}
{\scriptstyle \sum_{\Box'}}
(\mbox{tr}\,{A}^\dagger{\dot A}T_{h{\scriptscriptstyle\Box'}})^2,
\end{equation}
where ${\tilde z}$ measures dimensionless length along the flux tube
and the planes $\Box'$ intersect $\Box$ in single lines.
Our numerical solutions, modified by rotational
back reaction, give for the integrals
$ \tilde{\rho} \approx 5 F_\pi/e,
\ {\Lambda}_{\Box} \approx 40 / e^3 \, F_\pi,$ and $
{\Lambda}_{\Box'} \approx 16/ e^3 \, F_\pi$. The moments of inertia
${\Lambda}_{\Box}$ and
${\Lambda}_{\Box'}$
vary little with back reaction for quantum
excited states; however, the tension $\tilde{\rho}$ grows
linearly with $\omega$, due to compression of the
rotating flux tube.

To quantize this Lagrangian, we must rewrite the time derivatives in
eq.~(\ref{MQSkyrme}) in terms of canonical momenta. Under an $SO({N_F})$
transformation $h$, $\Sigma \rightarrow h\Sigma
h^T$, so that $A \rightarrow h A$ from eq.~(\ref{defSigt}). The
generators ${\cal I}_h$ of $H$ rotations thus generate
left-transformations of $A$. Similarly, we define body-centered
generators ${\cal I}_h'$ generating right
transformations, $A\rightarrow A h$. ${\cal
  I}_h$ and ${\cal I}_h'$ are related by an orthogonal transformation,
\begin{equation}
{\cal I}_h = R_{h h'} {\cal I}_{h'}',\qquad R_{hh'}=2 \,\,\mbox{tr}\,
\left(A^\dagger T_h A T_{h'} \right) \ .
\end{equation}
Noether's theorem then gives for ${\cal I}_h'$,
\begin{equation}
{\cal I}_h' = i \Lambda_h \,\,\mbox{tr}\, ({A}^\dagger \dot{A} T_h) \quad
\mbox{(no sum).}
\end{equation}
The flux tube solution $\Sigma(r,\theta)$ is invariant under the
action of ${\cal I}_{\Box}'+J_z$, so that ${\cal I}_\Box'$ is $-J_z$
(much like the Skyrme model, where $\vec I$ in the body
centered frame is $-\vec J$). It is also invariant under
$SO({N_F}-2)\subset SO({N_F})$, acting on the subspace $\Box_\bot$
orthogonal to $\Box$. This restricts $\Sigma$ to quantum states ${\cal I}'$
containing a singlet under $SO({N_F}-2)$.

Given these constraints, we write the Hamiltonian density
obtained from eq.~(\ref{MQSkyrme}) in terms of the physically observable
Noether charges, ${\cal I}_h$ and $J$,
\begin{equation}
\label{finalH}
H (\tilde{z}, t) =\tilde{\rho}+ {1\over{2{\Lambda}_{\Box}}} J_{z}^2
+{1\over {2{\Lambda}_{\Box'}}} (\ {\cal I}^2 - J_{z}^2\ ) \ ,
\end{equation}
which is quantized subject to the constraint ${\cal
I}_{\Box}'= -J_z$, with bosonic defects for ${N_F} > 3$.
For three flavors, this gives a spectrum of
states $(I,J_z)$ where $I$ and $|J_z|$ are
integers, $-I\le J_z \le I$, and $ {\cal I}^2 = I(I+1)$.

\centerline{{\bf Flux Tube Interactions}}

Our sigma model has both flux tube and skyrmion solutions. We show
that flux tube solutions are Alice strings. Moreover, loops of flux
tube can carry skyrmion number, just as loops of gauged Alice string
support monopole charge. We construct the skyrmions and discuss their
interactions with flux tubes, showing that only the topologically
trivial combination of two flux tubes can end on a skyrmion. This
suggests that, while baryons (i.e. skyrmions) are not confined, the
spinor sources which combine to form them are, with confinement
mediated by $Z_2$ flux tubes joining them.

First we consider twisted flux tubes. Our flux tubes have the Alice
property: some unbroken symmetries preserve a local vev but cannot be
extended globally, since transport around the string makes them
multivalued \cite{Alice}. This property seems ambiguous, as the flux
tube eq.~(\ref{Sigfinform}) allows many choices of underlying
$g(r,\theta)$. Double-valued choices --- such as $g(r,\theta) =
h(\theta)\,\, b(F(r)/2)$ --- give unbroken generators $g\,\,
T_{h\Box'} \,\, g^{-1}$ double-valued in $\theta$; while single-valued
choices --- such as $g(r,\theta) = h(\theta)\,\, b(F(r)/2)\,\,
h(\theta)$ --- leave all generators single-valued.  This ambiguity
persists even when we recast the flavor-dependent quark mass $\Sigma$
as an interaction between flavor gauge fields $g^{-1} (r,\theta)
\,\,\partial_\mu \,\, g(r,\theta)$ and shifted quarks $Q_{L} =
g(r,\theta) \,\, q_{L}.$ \cite{Balach} This identifies the flux tube
with a gauged string, with Wilson loop $ U(2\pi) = {\rm P}\exp{ \left
  ( \oint\vec{A} \cdot d\vec{l}\ \right ) }.$ However, the ambiguity
persists: $g(r,\theta)$ double-valued gives Wilson loop $U(2\pi)= {
  {\mbox{1\hskip-0.22em\relax l}}\,} - 2 \,\,{
  {\mbox{1\hskip-0.22em\relax l}}\,}_\Box$, making both quarks $Q_{L}$
and generators $T_{h\Box'}$ double-valued.  Single-valued
$g(r,\theta)$ produces instead single-valued quarks with an
$r$-dependent Wilson loop, giving $U(2\pi)= {
  {\mbox{1\hskip-0.22em\relax l}}\,}$ and single-valued generators at
$r=\infty$, but nontrivial $U(2\pi)$ and multi-valued generators
$T_{h\Box'}$ at finite $r$.

This ambiguity can be resolved physically, by considering adiabatic
transport of quarks around the flux tube. \cite{wilczek} Under such
transport, quarks remain in their mass eigenstates. These are fixed by
two terms: a flavor-independent bare Majorana mass $M$ (breaking
$SU(N_f) \rightarrow SO(N_f)$ explicitly) and the flavor-dependent
Majorana mass $\mu\Sigma$, where $\mu$ is the $SU(N_f)$-breaking vev.
These give mass eigenstates double-valued in $\theta$.  However, the
states have mass splitting $\Delta m^2 = -2i M (\mu - \mu^*)
\sin(F/2)$, with $M$ real. Thus the quarks are degenerate, and
unaffected by transport around the string, unless the vev $\mu$
misaligns in phase with the bare mass $M$. In that case, quarks at
finite radius have double-valued wave functions and Aharonov-Bohm
scatter off the flux tube, flipping the sign of their $T_{h\Box'}$
charges.~\footnote{We treat quarks with trajectories well-separated
  from the origin. This holds asymptotically inside the flux tube,
  where $m_i^{-1} < r < m_\pi^{-1}$ , for quark mass eigenvalues $m_i$
  .} Thus the flux tube acts as an Alice string in its core. This
analysis readily generalizes to other $\pi_2$-strings.  This
suggests the promise of observing similar effects in nonabelian physical
systems with global $\pi_2$-strings, such
as liquid He-$3a$.\cite{salmin}

Since twisted Alice loops can support monopole charge, we calculate
the skyrmion number of a twisted flux loop, $ \Sigma(z, r, \theta) =
A(z) \ \Sigma(r,\theta)\ A^{-1} (z) $ with $A(z) = \exp{(i z\, l\, n_h
  T_h)}.$ $2\pi$-periodicity implies that $l$ is integral for planar
$n_h=n_\Box$, and even otherwise. Thus any twisted flux loop can
deform to the planar flux loop $ \Sigma(z, r, \theta) = h(\theta + lz)
\ b(F(r)) \ h^{-1} (\theta + lz)$, taking values
within the subspace $SU(2)/SO(2) \sim S^2.$ Thus we can identify its
$\pi_3$ index with the map's Hopf number, {\em i.e.} the linking
number between any two fibers of constant $\Sigma$ in physical space. From
\cite{wuetc}, this linking number is precisely $l$ ---
the number of times a nontrivial fiber $\Sigma_0$ twists around the
loop's core, which has $\Sigma= -{ {\mbox{1\hskip-0.22em\relax
      l}}\,}_\Box$. Thus flux loops with an $l=1$ planar twist form
fundamental skyrmions; flux loops with nonplanar twist are trivial for
$N_f > 3$.

A nicer parametrization of the skyrmion stems from the exact sequence
\[
\pi_3\left( SO(N_f)\right)  \ \  \rightarrow \ \
\pi_3\left( SU(N_f)\right)  \ \  \rightarrow \ \
\pi_3\left( SU(N_f)/SO(N_f)\right)\ \  \rightarrow \ \
\pi_2\left( SO(N_f)\right) = 0 \  .
\]
This identifies skyrmions on the vacuum manifold with images of
skyrmions in $SU(N_f),$
$g(r, \hat{n} ) = \exp{(iF_s (r) \ \hat{n}_i\,\, T_{i \Box} )}$.
(Here $r$ and $\hat{n}$ are the radius and unit direction
vector in 3-space, and $F_s (r)$ approaches
$2\pi$ at $r=0$ and zero at $r=\infty$.) This gives an
axisymmetric skyrmion
\begin{eqnarray}
\label{skyrmion}
\lefteqn{\Sigma_s = { {\mbox{1\hskip-0.22em\relax l}}\,} + (\cos F_s - 1)\,\,
(1 - n_z^2) \,\,{ {\mbox{1\hskip-0.22em\relax l}}\,}_\Box
           } \nonumber\\
&& \quad + i\left(
\sin F_s \left(n_y \ \tau_{x\Box} + n_x \ \tau_{z\Box} \right)
+ 2\sin^2(F_s/2) \,\,n_z
\left(-n_x \ \tau_{x\Box} + n_y \ \tau_{z\Box} \right)
 \right)\ ,
\end{eqnarray}
after global spatial rotation.
Equation (\ref{Sigrexp}) for the flux tube lets us
identify the angular winding of a flux
tube with that of a skyrmion in the $xy$-plane ($n_z=0$). Thus, if their
radial forms coincided, we could deform the skyrmion's lower
hemisphere into a flux tube.  However, since
both $F$ and $F_s$ vary from $0$ to $2\pi$, the
skyrmion cannot end in a single flux tube.
Instead it joins only to flux tubes
where $F(r)$ ranges from $0$ to $4\pi$ --- that is,
configurations with two flux tubes, which are trivial.
Thus skyrmions are not confined.

However, objects which combine to form skyrmions can
interact with the flux tubes. Such ``half-skyrmions'' could arise as
external spinor sources in the underlying theory. They are
confined, as fundamentals cannot screen them. As mappings on $G/H$,
they appear precisely as half-skyrmions ---
objects of form (\ref{skyrmion}) with $F_s(r)$ ranging from $0$ to
$\pi$. Such objects are not defects in the conventional sense, since
they have linearly divergent energy --- like an unscreened point
source.  They can join to their
opposite winding counterparts via single flux tubes, restricting
linearly divergent energy to their separation distance.
Thus confinement of sources in an
$SO(N_c)$ gauge theory persists
in the low energy Skyrme model.

\acknowledgements
We thank Sidney Coleman, Glenn Boyd, Tom Imbo, and David Kaplan for
helpful conversations.  We thank our sponsors: Department of Energy
grant ~DOE-FG03-90ER40546,
Presidential Young Investigator Program grant ~PHY-8958081,
and the UC President's Postdoctoral
Fellowship program. KB and AM thank the Aspen Center for Physics
for hospitality during our final writing.

\begin{figure}
\caption{
  a.) Flux tube solutions $F(r)$ for $\lambda$ values $0.2,0.236,1.0,$
  and $2.0$. The dotted line corresponds to the physical value
  $\lambda_0=0.236$. Core size shrinks with increasing $\lambda$.
  \quad b.)  The above flux tubes' energy density $\rho$ ($\rho_0 +
  \rho_m$, from text).  Tension grows with $\lambda$.  }
\label{one}
\end{figure}

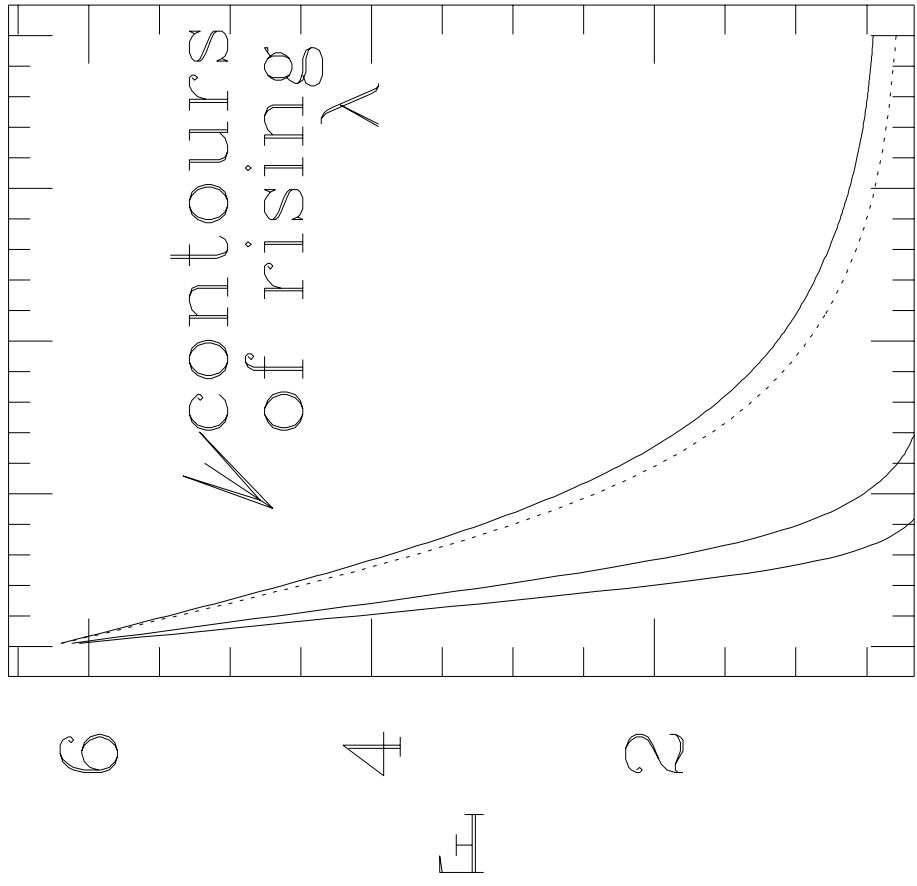
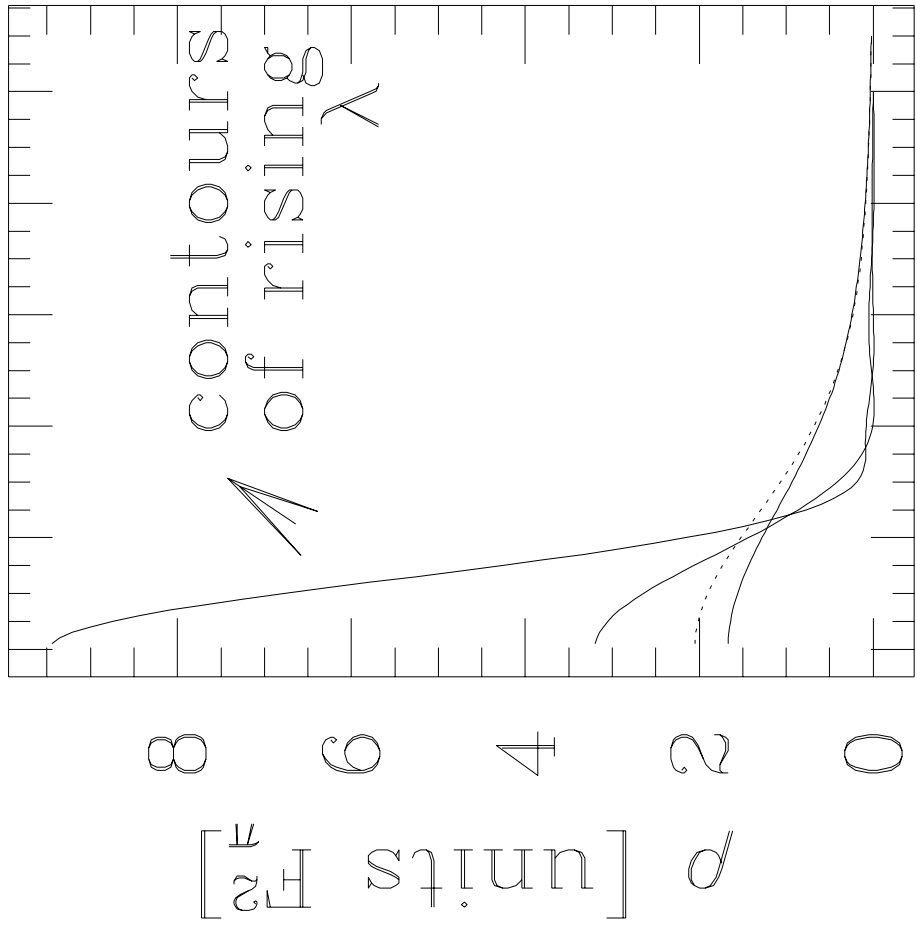
\end{document}